\documentclass[prl,aps,twocolumn,showpacs]{revtex4}
\usepackage{graphicx}
\begin{document}
\title{Self-organization of frozen light in near-zero-index media with cubic nonlinearity}
\author{A. Marini$^{1,}$}
\email{andrea.marini@icfo.es}
\author{F.~J.~Garc\'{\i}a~de~Abajo$^{1,2}$}
\affiliation{$^1$ICFO-Institut de Ciencies Fotoniques, Mediterranean Technology Park, 08860 Castelldefels (Barcelona), Spain}
\affiliation{$^2$ICREA-Instituci\'o Catalana de Recerca i Estudis Avan\c{c}ats, Barcelona, Spain}
\date{\today}
\begin{abstract}
We theoretically demonstrate the existence of frozen light in near-zero-index media with cubic nonlinearity. Light is stopped 
to a standstill owing to the divergent wavelength and the vanishing group velocity, effectively rendering, through nonlinearity, a 
positive-epsilon trapping cavity carved in an otherwise slightly-negative-epsilon medium.
By numerically solving Maxwell's equations, we find a soliton-like family of still azimuthal doughnuts, which we further study 
through an adiabatic perturbative theory that describes soliton evaporation in lossy media or 
condensation in actively pumped materials. Our results suggest applications in optical data processing and storage, quantum 
optical memories, and soliton-based lasers without cavities. Additionally, near-zero-index conditions can also be found in the 
interplanetary medium and in the atmosphere, where we provide an alternative explanation to the rare phenomenon of ball-lightning. 
\end{abstract}
\maketitle

\paragraph{Introduction --} 

Optical beams are generally unbound in free-space and in bulk media. Guidance and full spatial confinement of light are usually 
achieved by means of waveguides, mirrors, resonators, and photonic crystals. Alternatively, nonlinear self-organization can be 
exploited to compensate for diffraction of optical beams or dispersive broadening of pulses, enabling the formation of spatial 
and temporal solitons, respectively \cite{TrilloBook,Agrawal2001,Kivshar2003,ChSgChr2012}. Spatial self-trapping occurs in 
several optical systems, including photorefractive media \cite{EugenioOL1998,PTSolsOL2009}, liquid crystals 
\cite{ContiPRL2004,PecciantiNat2004}, and metamaterials \cite{ShadrivovJOptA2005,DongJPB2013}.
Remarkably, nonlinearity can act simultaneously on temporal and spatial domains to compensate for both diffraction and dispersion, 
thus enabling the formation of light bullets, spatio-temporal doughnuts, and X-shaped waves 
\cite{MihalachePRE2000,ContiPRL2003,MaMiWiTo2005,TornerOL2009,MinardiPRL2010}. 

Physical systems enabling either slow or fast light \cite{TsakmakidisNatLett2007,BoydJModPhys2009,KimOE2012} naturally enhance 
radiation-matter interaction, thus boosting nonlinear processes that can be efficiently used for active light control 
\cite{VlasovNature2005}, all-optical switching, and modulation \cite{MingaleevPRE2006,BajcsyPRL2009}. In particular, 
near-zero-index (NZI) media can slow down light propagation \cite{CiattoniPRA2013,JacobACSPhot2015} and enable extreme nonlinear 
dynamics \cite{CiattoniPRA2010}, enhanced second and third harmonic generation \cite{VincentiPRA2011}, active control of 
tunneling \cite{PowellPRB2009}, optical switching, and bistable response \cite{ArgyropoulosPRBB2012}. These materials 
naturally exist in nature, for example plasmas, transparent conductors, and metals near the their bulk plasma frequency 
$\omega_{\rm p}$ \cite{RaetherBook}. Besides, they can be artificially realized as 
waveguides close to modal cutoff \cite{VesseurPRL2013}, using surface phonon polaritons in GaAs quantum wells \cite{VassantPRL2012}, 
or by engineering subwavelength metallic nanowires, nano-spheres, or nano-circuits embedded in dielectric matrices. The latter 
strategy has enabled the development of epsilon-near-zero (ENZ) metamaterials, which have been investigated for applications 
such as enhanced transmission \cite{AluIEEE2006}, cloaking \cite{AluPRE2005}, energy squeezing in narrow 
channels \cite{SilveirinhaPRL2006}, and subwavelength imaging \cite{AluPRB2007,CastaldiPRB2012}. The ENZ regime is inevitably 
associated with high dispersion and is therefore accompanied by absorption, which can be suppressed by embedding externally 
pumped active inclusions in the NZI medium \cite{RizzaAPL2011}.  

In this Letter, we investigate self-organization of light in NZI media with Kerr-like instantaneous nonlinearity. 
In particular, we reveal the existence of fully confined doughnut-shaped solitons with vanishing Poynting vector and angular 
momentum. In practice nonlinearity enables digging a three-dimensional cavity for light, which in turn remains frozen and 
self-trapped. We study the effect of loss on stationary light doughnuts by developing a fully numerical soliton perturbative 
theory, finding that they evaporate over time due to inelastic absorption: their amplitude decreases, their frequency 
blueshifts slightly, and their radius increases. Conversely, if externally pumped active inclusions with inversion of 
population are embedded within the NZI medium, the opposite scenario takes place and azimuthal doughnuts condensate over time. 
These findings demonstrate the possibility to freeze light beams in ENZ media, with potential applications in optical data 
processing and storage, quantum optical memories, and NZI lasers operating without cavities. While confinement in random lasers 
is generally brought by Anderson localization \cite{WiersmaNatPhys2008,ContiPRL2008,SegevNatPhoton2013}, in the case of NZI 
media self-trapping is provided by optical nonlinearity in the form of solitary waves. Interestingly, ENZ conditions are found 
also in the interplanetary medium and in the atmosphere, and we argue that our theoretical results may provide insight 
into ball-lightning (BL) formation \cite{HandelJGRA1994,AbrahamsonNatLett2000,CenPRL2014}.

\paragraph{Model --} We consider a generic NZI medium with Drude temporal response and 
instantaneous Kerr-like nonlinearity. Both of these ingredients ensue from free-particle temporal dynamics,
which is characteristic of plasmas, metals, transparent conductors, and ENZ metamaterials, all examples 
of NZI media. In particular, Kerr-like nonlinearity naturally arises from the ponderomotive force in plasmas 
and metals \cite{GinzburgOL2010}, and is well represented by the constitutive relation between the displacement vector 
${\rm Re}[{\cal D}(t)]$ and the electric field ${\rm Re}[{\cal E}(t)]$: 
\begin{eqnarray}
{\cal D}(t) & = & \epsilon_0 \int_0^{\infty} \epsilon(t') {\cal E}(t - t') d t' + \label{DSPLVECEQ} \\
            &   & \epsilon_0 \chi_3\left\{|{\cal E}(t)|^2{\cal E}(t) + (1/2) \left[ {\cal E}(t) \cdot {\cal E}(t) \right] {\cal E}^*(t) \right\}, \nonumber
\end{eqnarray}

\begin{figure}[t]
\centering
\begin{center}
\includegraphics[width=0.4\textwidth]{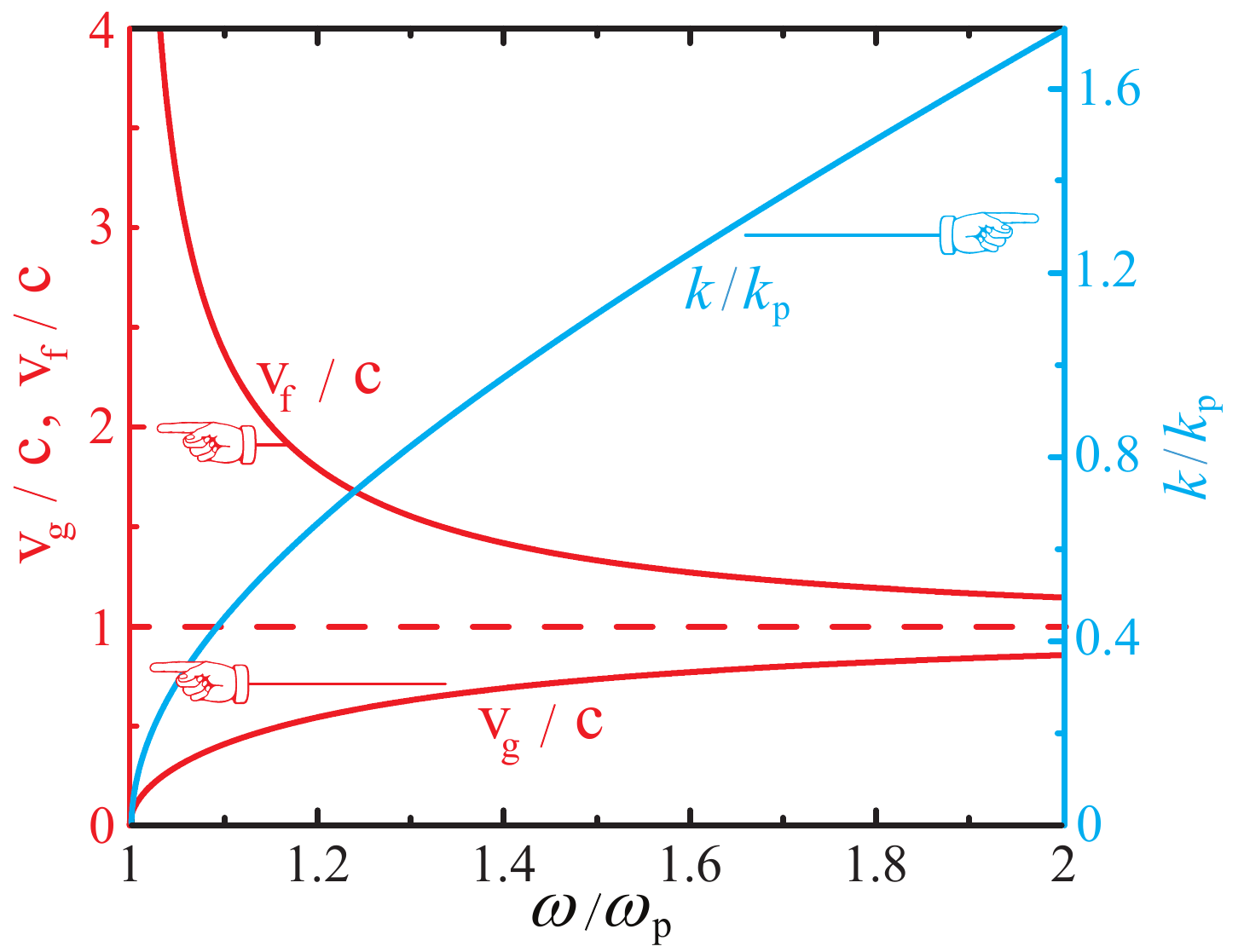}
\caption{(Color online) Dispersion relation $k(\omega)$ (cyan right $y$-axis) and phase and group velocities (${\rm v}_f$ and ${\rm v}_g$, 
red left $y$-axis) of TEM waves in the linear loss-less limit ($\chi_3,\gamma\rightarrow 0$). All quantities are plotted 
in dimensionless units: the angular frequency $\omega$ is normalized to the plasma frequency $\omega_{\rm p}$, the wave-vector 
$k$ is normalized to $k_{\rm p} = \omega_{\rm p}/c$, and the phase and group velocities are normalized to the speed of light in vacuum $c$. 
The red dashed line indicates the dispersion-less limit ${\rm v}_f = {\rm v}_g = c$.}
\label{Fig1}
\end{center}
\end{figure}
 
\noindent where $\epsilon_0$ is the vacuum permittivity, $\chi_3$ is the nonlinear susceptibility of the
medium, $\epsilon (\tau) = \delta(\tau) + \omega_{\rm p}^2(1-e^{-\gamma \tau})/\gamma$ is the Drude temporal 
response function, $\delta(\tau)$ is the Dirac delta-function, $\omega_{\rm p}$ is the plasma frequency, and 
$\gamma$ is the temporal damping rate due to inelastic collisions. Optical propagation is governed by the wave equation 
\begin{equation}
\nabla\times\nabla\times{\cal E} = - \mu_0 \partial_t^2 {\cal D}, \label{DBLCRLEQ}
\end{equation}
where $\mu_0$ is the vacuum permeability. In the linear limit $\chi_3\rightarrow 0$, homogeneous transverse 
electromagnetic (TEM) waves are solutions of Eq. (\ref{DBLCRLEQ}) given by 
${\cal E} =  {\bf e}_0 e^{i{\bf k}\cdot{\bf r} -i\omega t}$, 
where ${\bf k}\cdot{\bf e}_0 =0$. The angular frequency $\omega$ and the wave-vector ${\bf k}$ satisfy the 
dispersion relation $k(\omega) = (\omega/c)\sqrt{\epsilon(\omega)}$, where $c$ is the speed of light in free space and
$\epsilon(\omega) = 1 - \omega_{\rm p}^2/\omega(\omega+i\gamma)$ is the frequency-dependent dielectric constant, which is
given by the Fourier transform of the Drude temporal response function $\epsilon(\tau)$. The linear dispersion relation
of TEM waves $k(\omega)$ is depicted in Fig. \ref{Fig1} in the lossless limit $\gamma \rightarrow 0$, together with the 
phase and group velocities ${\rm v}_f(\omega) = \omega/k(\omega),~{\rm v}_g(\omega) = {\rm d}\omega/{\rm d}k$. Note the 
cutoff of TEM waves at the plasma frequency $\omega = \omega_{\rm p}$, where the medium enters the ENZ regime, the phase 
velocity diverges, and the group velocity vanishes \cite{CiattoniPRA2013,JacobACSPhot2015}. 

\begin{figure}[t]
\centering
\begin{center}
\includegraphics[width=0.475\textwidth]{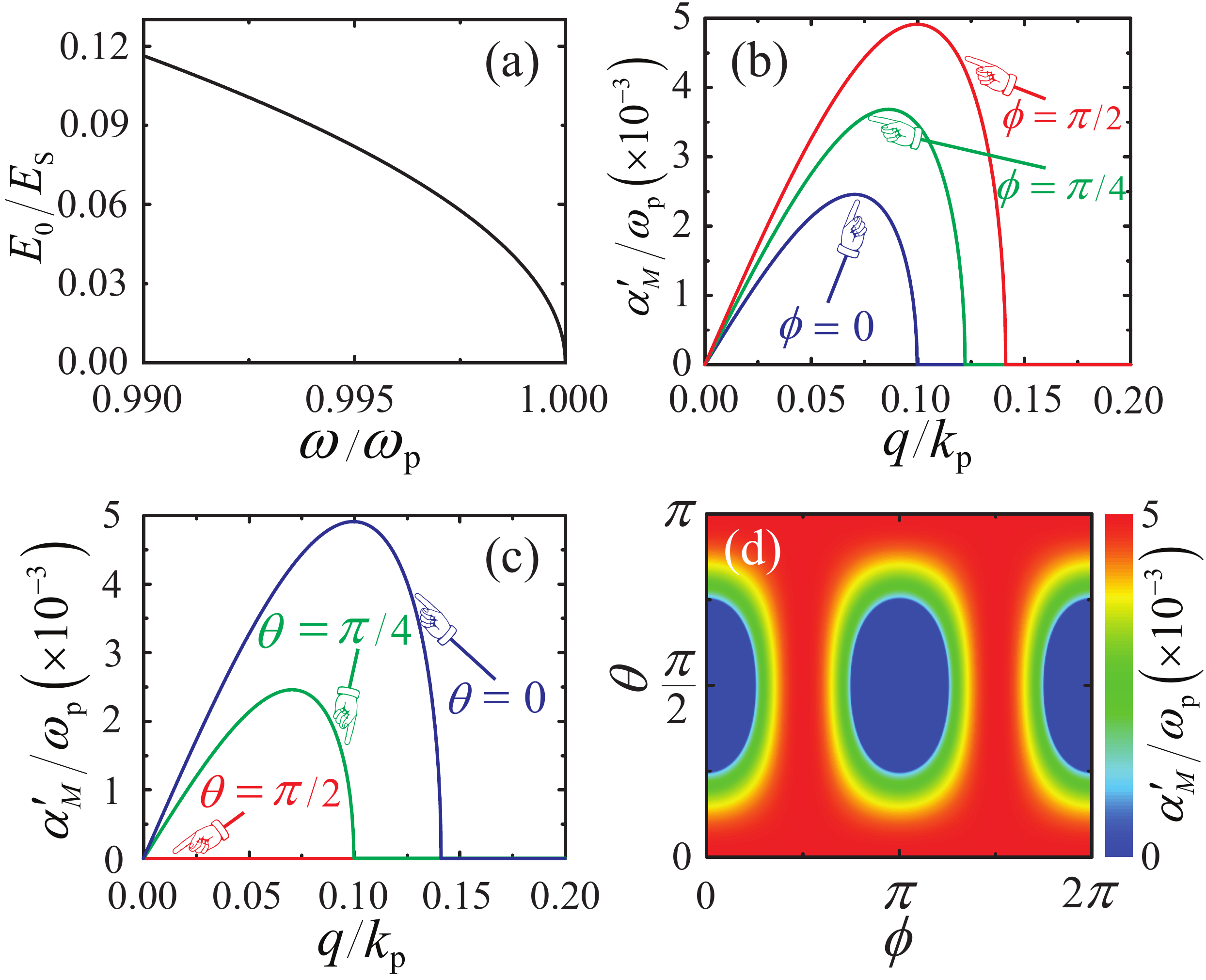}
\caption{(Color online) {\bf (a)} Nonlinear dispersion of zero-index homogeneous modes existing for angular frequencies $\omega$ 
smaller than the cutoff $\omega_{\rm p}$. The electric field amplitude $E_0$ is normalized to the
scaling electric field $E_{\rm S} = \chi_3^{-1/2}$. {\bf (b,c)} Maximum instability growth $\alpha'_{\rm M}$ (normalized to the 
plasma frequency $\omega_{\rm p}$) as a function of the perturbing wave-vector modulus $q$ (normalized to $k_{\rm p}=\omega_{\rm p}/c$) 
for several directions in the reciprocal space: {\bf (b)} $\theta = \pi/4$, $\phi = 0, \pi/4,\pi/2$, and {\bf (c)} $\phi = 0$, and 
$\theta = 0, \pi/4,\pi/2$. {\bf (d)} Contour-plot of the maximum instability growth $\alpha'_{\rm M}$ (normalized to the plasma frequency 
$\omega_{\rm p}$) as a function of $\theta,\phi$ for a fixed perturbing wave-vector modulus $q/k_{\rm p} = 0.1$.}
\label{Fig2}
\end{center}
\end{figure}

\paragraph{Homogeneous nonlinear modes --}

Owing to the vanishing group velocity, nonlinear effects are dramatically enhanced in the ENZ regime \cite{CiattoniPRA2010,VincentiPRA2011}. 
For $\omega < \omega_{\rm p}$, homogeneous modes with vanishing wavenumber, infinite phase velocity, and zero group velocity can be found by 
neglecting damping and setting ${\cal E} = {\bf E}_0 e^{-i\omega t}$, with $E_0 = \sqrt{-2\epsilon(\omega)/(3\chi_3)}$. 
The resulting dispersion relation is plotted in Fig. \ref{Fig2}(a). We find zero-index homogeneous modes to have a cutoff at the plasma 
frequency $\omega_{\rm p}$, where the electric field amplitude drops to zero. In order to evaluate the stability of homogeneous modes, we 
perturb them with small-amplitude waves: 
${\cal E} =  \left[ {\bf E}_0 + \delta{\bf E}_1 e^{i{\bf q}\cdot{\bf r} + \alpha t} + \delta{\bf E}_2^* e^{-i{\bf q}\cdot{\bf r} + \alpha^* t} \right] e^{-i\omega t}$, where $\delta{\bf E}_1,\delta{\bf E}_2$ are the perturbation amplitudes with wave-vector ${\bf q}$
and temporal growth eigenvalue $\alpha$. Inserting this expression in Eqs. (\ref{DSPLVECEQ}) and (\ref{DBLCRLEQ}) and retaining only the 
lowest-order terms in $\delta{\bf E}_1$ and $\delta{\bf E}_2$, we find a homogeneous system of linear equations, whose non-trivial solutions 
are signaled by the vanishing of the secular determinant \cite{EPAPS}. This condition determines the complex temporal eigenvalues $\alpha$. 
Instabilities are then associated with positive real parts of the eigenvalue $\alpha$, indicating unbound amplification of the perturbation. 
We plot results of the stability analysis in Figs \ref{Fig2}(b-d), and in particular, we depict the maximum of the real part of the 
eigenvalue, $\alpha'_{\rm M}$. In analogy to standard modulation instability in $1$D paraxial systems \cite{Kivshar2003}, the gain spectrum 
of the perturbations is non-vanishing within a finite wavevector window and is peaked at a characteristic wavevector modulus. However, 
in contrast to $1$D paraxial systems, the gain spectrum is $3$D and has a non-trivial dependence on polar and azimuthal angles 
($\theta$,$\phi$) of the perturbation wavevector ${\bf q}$.

\paragraph{Still azimuthal doughnuts --}

\begin{figure}[t]
\centering
\begin{center}
\includegraphics[width=0.45\textwidth]{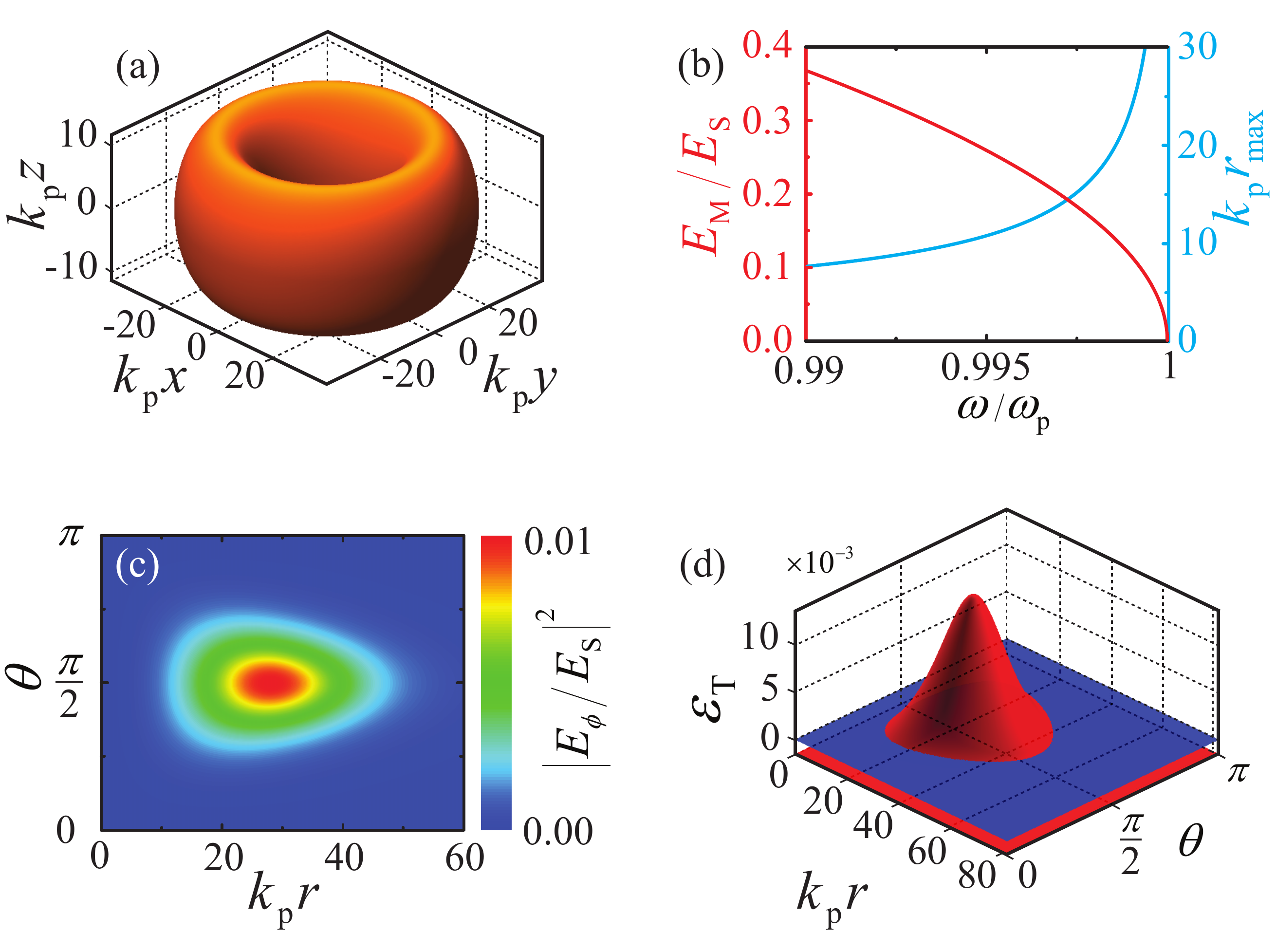}
\caption{(Color online) {\bf (a)} Iso-surface $|E_{\phi}({\bf r})/E_{\rm S}|^2 = 0.005$ of a still doughnut with maximum squared amplitude 
$|E_{\rm M}/E_{\rm S}|^2 = 0.01$, where $E_{\rm S}=1/\sqrt{\chi_3}$ is the scaling field amplitude, excited at an angular frequency 
$\omega/\omega_{\rm p} =  0.9992$. {\bf (b)} Soliton maximum amplitude $E_{\rm M}/E_{\rm S}$ (red left $y$-axis) and its corresponding radius 
$r_{\rm max}$ (cyan right $y$-axis) as a function of angular frequency $\omega/\omega_{\rm p}$.
{\bf (c)} Contour-plot in the $r$-$\theta$ plane of the dimensionless intensity profile $|E_{\phi}(r,\theta)/E_{\rm S}|^2$ and 
{\bf (d)} total dielectric permittivity profile $\epsilon_{\rm T}(r,\theta)$ (red surface) associated with the still doughnut of (a). 
The blue plane in (d) represents the metal-dielectric transition plane $\epsilon_{\rm T}(r,\theta) = 0$. All quantities 
are plotted in dimensionless units: the angular frequency $\omega$ is normalized to the plasma frequency $\omega_{\rm p}$, while spatial 
coordinates are normalized to the inverse of the plasma wave-vector $k_{\rm p}^{-1}$. }
\label{Fig3}
\end{center}
\end{figure}

The modulation instability scenario strongly suggests the presence of still $3$D solitons 
in NZI media. In order to verify this hypothesis, we transform Eq. (\ref{DBLCRLEQ}) into spherical coordinates and search for 
azimuthally-polarized solutions: ${\cal E} = E_{\phi} (r,\theta) e^{-i\omega t} \hat{\phi}$. 
As Eq. (\ref{DBLCRLEQ}) is invariant under a constant phase shift, without any loss of generality 
we can assume that the electric field envelope is real $E_{\phi} (r,\theta)\in \Re$, meaning that we are seeking non-propagating 
solutions which are not accompanied by a phase flow. Indeed, assuming that such solutions exist, we show that the Poynting 
vector vanishes thoroughly (see SI \cite{EPAPS}). Besides, we seek localized soliton-like solutions vanishing at $r\rightarrow\infty$ 
and at $r=0$, $\theta = 0,\pi$ owing to the azimuthal polarization. Upon examination of the asymptotical expansion of 
Eq. (\ref{DBLCRLEQ}) for $r\rightarrow\infty$, we find that $3$D soliton-like azimuthal solutions can actually exist only in the 
ENZ regime (see SI \cite{EPAPS}). Thus, we discretize derivatives with respect to the radius $r_n$ and the polar angle $\theta_m$ 
and then transform the differential wave equation for the electric field into a nonlinear algebraic system for the electric 
field amplitudes $E_{\phi,n,m}$ in the two-dimensional grid $r_n,\theta_m$ (see SI \cite{EPAPS}). We solve this 
nonlinear algebraic system by means of an iterative Newton-Raphson algorithm, and find a family of still azimuthal doughnuts 
[see Fig. \ref{Fig3}(a)] for $\omega < \omega_{\rm p}$, which presents a cutoff at $\omega_{\rm p}$, where the soliton loses 
localization and its amplitude vanishes. The frequency-dependent maximum amplitude and the corresponding radius of the still doughnut 
family are plotted in Fig. \ref{Fig3}(b), while a $r$-$\theta$ contour-plot of the squared electric field profile 
$|E_{\phi}({\bf r})/E_{\rm S}|^2$ (normalized to the scaling field $E_{\rm S} = 1/\sqrt{\chi_3}$) of the still doughnut at 
$\omega/\omega_{\rm p} = 0.995$ is depicted in Fig. \ref{Fig3}(c). The total dielectric permittivity profile 
$\epsilon_{\rm T} (\omega,{\bf r}) = \epsilon (\omega) + (3/2)\chi_3|E_{\phi}(\omega,{\bf r})|^2$ is shown in Fig. \ref{Fig3}(d). 
Importantly, in the soliton existence domain $\omega < \omega_{\rm p}$, the linear dielectric constant is negative $\epsilon(\omega) < 0$, 
and thus, at long radius where the electric field amplitude is small, the NZI medium is metal-like. Conversely, in the volume around the 
radius $r_{\rm max}$ for which the electric field is maximum, nonlinearity is non-negligible and the total dielectric permittivity 
is positive $\epsilon_{\rm T}({\bf r}_{\rm max}) >0$ (dielectric-like). From here we see that the existence of still azimuthal doughnuts 
originates in the extraordinary ability of nonlinearity to dig a dielectric-like $3$D cavity within a metal-like environment. This 
scenario is unique of NZI media, which prevent propagation of the fields outside the induced-dielectric trapping cavity. We emphasize 
that modulation instability enables the excitation of non-propagating solitons starting directly from unstable homogeneous waves with 
frequency falling in the ENZ regime.

\paragraph{Doughnut evaporization/condensation --}

\begin{figure}[t]
\centering
\begin{center}
\includegraphics[width=0.45\textwidth]{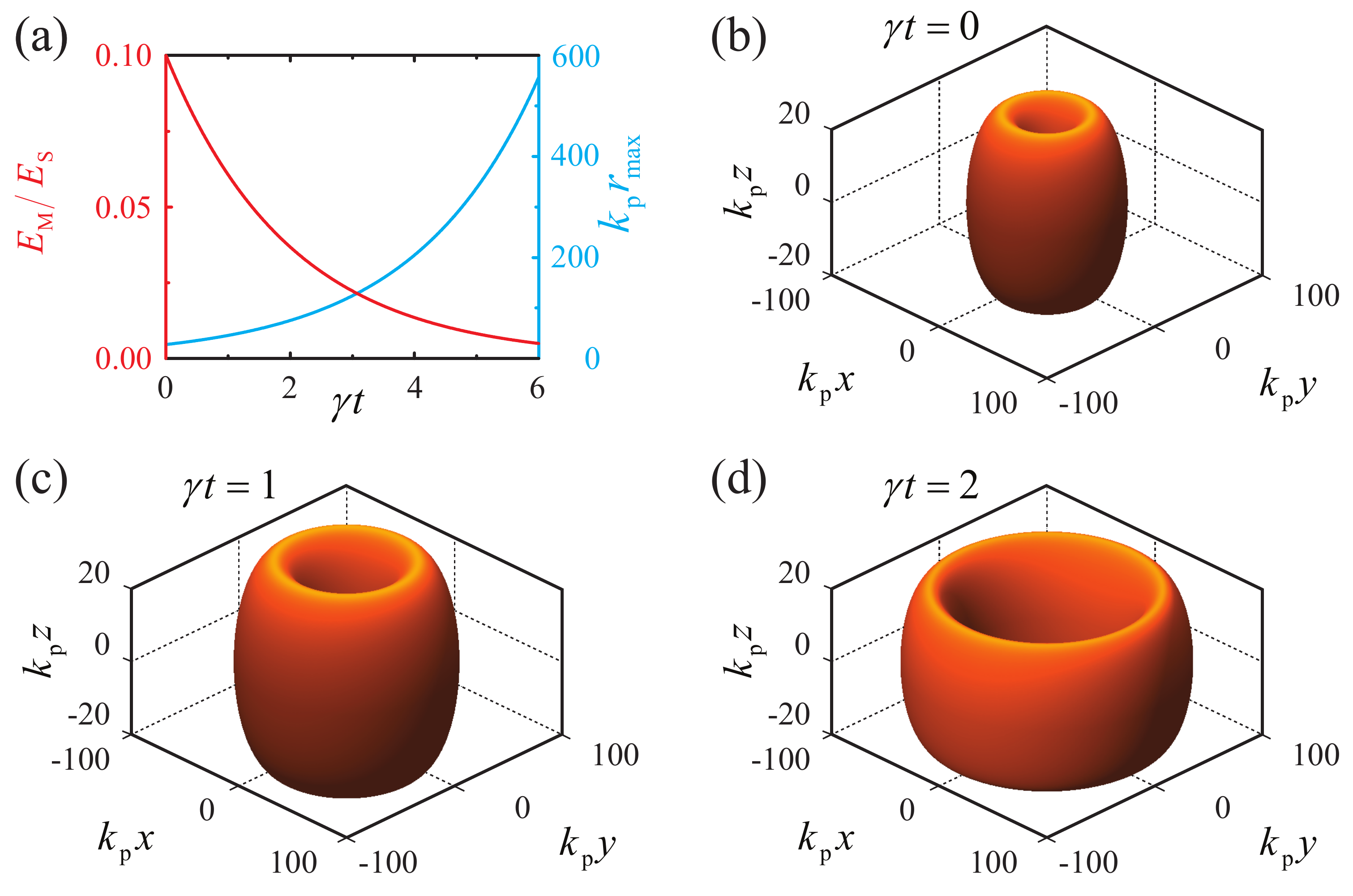}
\caption{(Color online) {\bf (a)} Soliton maximum amplitude $E_{\rm M}/E_{\rm S}$ (red left $y$-axis) and its corresponding 
dimensionless radius $k_{\rm p} r_{\rm max}$ (cyan right $y$-axis) as a function of dimensionless time $\gamma t$, where $\gamma$ is a 
phenomenological damping. {\bf (b-d)} Iso-surfaces $|E_{\phi}({\bf r})/E_{\rm S}|^2 = 0.001$ of the time-evolving doughnut with initial 
condition as in Fig. \ref{Fig3}(a) for {\bf (b)} $\gamma t = 0$, {\bf (c)} $\gamma t = 1$, and {\bf (d)} $\gamma t = 2$.}
\label{Fig4}
\end{center}
\end{figure}

In standard transparent media, the main quantity accounting for optical propagation is the Poynting vector, representing the photon flux 
per unit time. For our trapped solitons, the Poynting vector is thoroughly vanishing (see SI \cite{EPAPS}), so we describe doughnut 
self-trapping through the optical-cycle-averaged density of electromagnetic energy $u = (1/2){\cal E}\cdot{\cal D}$. Now, if absorption 
is taken into account, the energy density is expected to be damped and vanish exponentially over time. A numerical verification of this 
hypothesis could consist in temporally evolving Eq. (\ref{DBLCRLEQ}) with the doughnut initial condition. However, temporal evolution requires 
nonlinear $3$D finite-difference-time-domain (FDTD) numerical simulations, which are computationally demanding.
Besides, traditional approaches used in dielectric and plasmonic waveguides \cite{AfsharOE2009,MariniPRA2011,SkryabinJOSAB2011} 
relying on the slowly-varying-envelope approximation (SVEA) can not be used, as the SVEA does not hold in the ENZ regime \cite{CiattoniPRA2010}. 
Instead, we have developed a soliton perturbation theory (see SI \cite{EPAPS}) capable of accounting for both damping and amplification 
(e.g., in systems containing externally pumped active inclusions within the NZI medium) under the assumption that (i) damping ($\gamma>0$) 
or (ii) gain ($\gamma<0$) are much smaller than the soliton angular frequency $\omega$. We further assume that the temporal evolution of the 
still doughnut adiabatically follows the soliton family, finding that the soliton amplitude (i) decays or (ii) increases over time following 
the exponential law $E_{\rm M}(t) = E_0e^{-\gamma t/2}$, where $E_0$ is the initial field amplitude and $\gamma$ is a phenomenological 
absorption/pumping rate. Accordingly, the doughnut (i) expands and blueshifts or (ii) shrinks and redshifts in either case (see SI \cite{EPAPS}). 
The time-dependent field amplitude (blue left $y$-axis) and doughnut radius (red right $y$-axis) are plotted in Fig. \ref{Fig4}(a) for a 
representative example, along with three snap-shots of the iso-surface $|E_{\phi}({\bf r})/E_{\rm S}|^2 = 1\times10^{-3}$
at different times in Figs. \ref{Fig4}(b-d), where we have assumed as initial condition the doughnut of Fig. \ref{Fig3}(a) and (i) damping 
($\gamma>0$) (For the full temporal evolution see movie in the SI \cite{EPAPS}). The doughnut evaporates over time, as its amplitude decreases 
and its radius increases. The gain scenario (ii) ($\gamma<0$) can be interpreted by inverting the temporal direction, so that the doughnut 
condensates over time, as its amplitude increases and its radius decreases. 

\paragraph{Ball-lightnings? --} BLs are rare lightning events with hitherto unknown theoretical explanation 
\cite{HandelJGRA1994,AbrahamsonNatLett2000,CenPRL2014}. BLs emit broadband radiation and can either propagate or stand still. 
Initially considered as myth, BLs have puzzled scientists for centuries and their existence
has been questioned until the first recent experiment able to measure their spectrum \cite{CenPRL2014}. Understanding of the nature 
of BLs is still unsatisfactory as they can not be easily reproduced in laboratory. Among the several theories trying to explain 
their nature, the so-called maser-caviton theory \cite{HandelJGRA1994} suggests that BLs are localized high-field solitons forming 
a cavity surrounded by plasma. Indeed, during thunderstorms, atmosphere can get ionized and become a NZI medium with a plasma frequency 
falling in the terahertz-microwave spectral region, where rotational levels of water can be excited. The ensuing emitted 
radiation is thought to remain self-trapped and heat up the air, thus emitting broadband blackbody radiation \cite{HandelJGRA1994}. 
This theory explains some general aspects of BLs, but it does not provide any quantitative description of the self-induced soliton cavity. 
Following our rigorous calculations, we speculate that BLs may ensue from a self-organization process in the ENZ regime, where we 
observe the existence of still doughnut solitons, as discussed above. The actual spherical shape of BLs observed in experiments 
\cite{CenPRL2014} may be due to mixed polarization, higher order nonlinear effects, or the intrinsically incoherent nature of radiation 
emitted in the atmosphere. The ENZ condition would explain the infrequency of the phenomenon and provides an insightful signature for 
experimental investigations.

\paragraph{Conclusions --} Our investigation of self-organization phenomena in NZI media with cubic nonlinearity has resulted in the 
demonstration that zero-index nonlinear waves are unstable in all spatial directions and that still azimuthally polarized 
self-trapped doughnuts can be excited. We have discussed the existence domain of this $3$D soliton family with thoroughly vanishing 
Poynting vector and provided details on its characteristics. Besides, we have studied the effect of loss/amplification, 
finding that still light doughnuts evaporate/condensate over time, respectively. Our model applies to any NZI medium with cubic 
nonlinearity and our results are universal as they are rescaled to the relevant physical quantities (plasma frequency $\omega_{\rm p}$, 
plasma wave-vector $k_{\rm p}$, Kerr coefficient $\chi_3$) of any specific medium in this regime (e.g., metals, transparent conductors, 
plasmas, and metal-dielectric ENZ metamaterials). Our findings pave the way for the development of novel applications in optical 
data processing and storage, the realization of quantum optical memories, and the design of soliton-based lasers without 
cavities. Incidentally, NZI conditions can be found also in the interplanetary medium and in the atmosphere, and we have discussed 
possible relationships between our results and ball-lightning formation.

\noindent A.M. is supported by an ICFOnest$+$ Postdoctoral Fellowship (Marie Curie COFUND program). A.M. acknowledges fruitful discussions 
with Alessandro Ciattoni, Mario Raparelli, and Carlo Rizza.

\end{document}